\documentclass[prx,aps,twocolumn,floats,showpacs,final,groupedaddress]{revtex4-2}
\usepackage{leading}
\usepackage[latin3]{inputenc}
\usepackage[makeroom]{cancel}
\usepackage{graphicx}
\usepackage{amsmath}
\usepackage{amsfonts}
\usepackage{amssymb}

\usepackage{color}
\usepackage[left=2cm,right=2cm,top=2cm,bottom=2cm]{geometry}
\usepackage{graphicx} 
\usepackage{dcolumn}
\usepackage{bm}
\usepackage{simplewick}
\usepackage{array}
\usepackage{appendix}
 \usepackage {float}
\RequirePackage[
   hyperindex,colorlinks,bookmarksnumbered,
   plainpages=true,pdfstartview=FitH]{hyperref}
\hypersetup{linkcolor=blue,urlcolor=blue,citecolor=blue}
\usepackage{hyperref}
\usepackage{mathrsfs}
\definecolor{purple}{rgb}{0.5,0,0.6}

\usepackage{ulem}
\renewcommand{\emph}[1]{\textit{#1}}
\definecolor{darkblue}{rgb}{0,0,0.5}
\definecolor{darkgreen}{rgb}{0,0.5,0}
\definecolor{darkred}{rgb}{.7,0,0}
\definecolor{purple}{rgb}{0.5,0,0.6}
\definecolor{orange}{rgb}{1,0.5,0}
\definecolor{grey}{rgb}{.6,.6,.6}
\definecolor{lightpink}{rgb}{1,0.7,0.75}
\definecolor{pink}{rgb}{1,0.4,0.58}
\definecolor{deeppink}{rgb}{1,0.08,0.58}

\renewcommand{\emph}[1]{\textit{#1}}

\newcommand{\ngg}{N_{\rm g}}

\begin{document}

\title{Multi-mode Coulomb blockade oscillations}
\author{D. B. Karki}
\affiliation{Materials Science Division, Argonne National Laboratory, Argonne, Illinois 60439, USA}
\begin{abstract}
We develop a theory of Coulomb blockade oscillations in transport and thermodynamic properties of a mesoscopic device having multiple charging energy modes. This setup can be realized using a nanoelectronic circuit comprising coupled hybrid metal-semiconductor islands. We show that this device can have various distinctive operational regimes depending on the strength of charging modes and base temperature. We focus on three different regimes; quantum regime, thermal regime and quantum-thermal mixed regime, in which the shape of the Coulomb blockade oscillations manifests well-defined features that can be accessed via conductance measurements. Our theory covers full crossover among these regimes, and also accounts for an accidental vanishing of one of the charging modes.
\end{abstract}
\maketitle

\section{Introduction}
Interactions between electrons in low-dimensional systems give rise to a number of strongly correlated phenomena. Many of these effects can also be studied using well-characterized mesoscopic devices. A typical mesoscopic device is the single-electron transistor, which contains a small metallic grain tunnel coupled to two electronic reservoirs~\cite{Matveev_1991}. In such a confined structure, Coulomb interaction plays an important role, and a plethora of nontrivial effects emerges~\cite{Matveev_1991,AL, Averin1992, sett}. For example, a remarkable manifestation of electron interactions at low temperatures is the phenomenon of Coulomb blockade (CB), where electron transport is exponentially suppressed due to the large charging energy $E_c\gg T$ of the grain. 

Coulomb blockade can be lifted by applying voltage $V_{\rm g}$ to the additional central gate, which modifies the electrostatic energy of the grain as
\begin{equation}\label{sg1}
\mathbb{E}_e=E_c\left(Q-\ngg\right)^2,
\end{equation}
where $Q$ is the average charge of the grain, and the dimensionless parameter $\ngg$ is proportional to $V_{\rm g}$ (for simplicity of presentation, throughout the paper we set $e=\hbar=k_{\rm B}=1$). Therefore, for $\ngg=(n+1/2)$ the charge states $n$ and $(n+1)$ become degenerate, signaling the complete absence of CB effect. Consequently, the conductance measurement of the single-electron transistor shows periodic peaks, also known as CB oscillations, as a function of gate voltage. These oscillations exist as long as the barrier transmission is less than unity, i.e., the presence of charge granularity, and the precise form of their amplitudes depends on the relationship between $T$ and $E_c$. Unique features of CB oscillations in the quantum $E_c\gg T$, and the thermal $E_c\ll T$ regimes of the device have been thoroughly investigated~\cite{Flensberg_1993,Matveev1995, Furusaki1995b,Pierre_2016, td}.

Different aspects of CB have also been studied, both theoretically and experimentally, in a device consisting of coupled metal grains~\cite{dd1, kamdd, kamdd1, koven}. In a recent experiment~\cite{ccd}, such a setup has been realized using a nanoelectronic circuit comprising coupled hybrid metal-semiconductor islands, where coupling between the grains was achieved by connecting them through a short electronic channel implemented via a fully tunable quantum point contact (QPC). Remarkable predictions for a number of effects resulting from the competition between the screening of each individual grain charge, and the mediated charge coupling between the two grains have also been reported~\cite{ccd, KBM1}.

In the general case of a device consisting of the coupled grains, the Coulomb interactions within and between the grains play an important role. In addition, grains can be of different sizes and have different charging energies. We note that while the Coulomb interactions are typically fixed by the geometry of the device, the base temperature can in principle be an arbitrary parameter. Given these considerations, the obvious definition of quantum and thermal regimes defined earlier becomes more involved in the case of Coulomb coupled devices. The latter can be appreciated by a simple consideration of the typical charging energy Hamiltonian of the coupled CB structures
\begin{equation}\label{sg2}
\mathcal{H}=E_c^{(1)}\hat{N}_1^2+E_c^{(2)}\hat{N}_2^2+2E_I\hat{N}_1\hat{N}_2,
\end{equation}
where $E_I$ is the Coulomb interactions between the grains, $\hat{N}_i$ stands for the charge operator of the grain $i$, and we assumed the absence of gate voltages. $\hat{N}_i$ obviously depends largely on the details of the geometry of the device such as number of contacts, their transparencies and the way how the grains are being connected~\cite{kamdd}. Nevertheless, some crucial insights can also be gained by disregarding those details.

Considering the basis of charge operators $\hat{N}_i$, the eignevalues of the Hamiltonian~\eqref{sg2} takes the form $\lambda_\pm=\lambda_e\pm\sqrt{\lambda^2_0+E^2_I}$ with $\lambda_{e/o}=(E^{(1)}_c\pm E^{(2)}_c)/2$, i.e., eigenvalues consist of the light $\lambda_-$ and heavy $\lambda_+$ modes. Importantly, the light eigenmode $\lambda_-$ vanishes at the characteristics value of inter-grain interaction $E^*_I=(E_c^{(1)}E_c^{(2)})^{1/2}$. In this case, the coupled device formally becomes identical with the corresponding single grain setup with effective charging energy $E^{(1)}_c+E^{(2)}_c$, and the grain charge operator $\propto (\hat{N}_1+\hat{N}_2)$. This is a sharp phase transition given that the charging energy is only the interaction in the system. Although in these discussions we disregarded the detailed geometry of the device, it is arguably clear that any double-grain setups host two charging energy modes. Unless extra fine-tuning is applied, these modes can be resolved into the light mode $\mathbb{E}_{\rm l}$, and the heavy mode $\mathbb{E}_{\rm h}$ for the given value of interaction parameters. Such a situation also arises even in the absence of Coulomb coupling if the grains are of reasonably different sizes.

From these discussions, it is clear that the coupled-charge grain setup can have various CB regimes as different inequalities between the temperature and charging energy modes can be formed. In general, four unique CB regimes can be categorized: i) quantum regime $T\ll\left(\mathbb{E}_{\rm l}, \mathbb{E}_{\rm h}\right)$, ii) thermal regime $T\gg\left(\mathbb{E}_{\rm l}, \mathbb{E}_{\rm h}\right)$, iii) quantum-thermal mixed regime $\mathbb{E}_{\rm l}\ll T\ll \mathbb{E}_{\rm h}$, and iv) the regime of phase transition $\mathbb{E}_{\rm l}=0$. Therefore, for the actual comparison of CB oscillations measurements in double grain setups with theoretical predictions, one requires a theory with broad applicability without relying on the special inequality between temperature and charging energy modes. Here we develop such an approach.

Recently, a theoretical framework has been advanced for the study of transport properties in the quantum regime $T\ll\left(\mathbb{E}_{\rm l}, \mathbb{E}_{\rm h}\right)$ of double-charge quantum island setup~\cite{KBM}. The present theory takes into account both intra and inter-island Coulomb interactions, and is applicable in all four regimes discussed above. This work also aims to illustrate the importance of Coulomb coupling in realistic double-grain devices, and shed light on the possibility of observing interaction-dependent temperature scaling in the linear conductance of such a device. Exploring certain well-defined features of different asymptotic regimes discussed above is an additional intent of this work.

The paper is organized as follows. In Sec.~\ref{model} we introduce a bosonization-based model to describe the experimental setup consisting of double-charge quantum islands. The scattering theory of boson accounting for both intra and inter-island Coulomb interactions relevant to the double-grain devices is constructed in Sec.~\ref{sth}. Elaborated discussions about the charging energy modes and their implications are presented in Sec.~\ref{chm}. In Sec.~\ref{chargec} we outline the derivation of charge current. Section~\ref{lc} is devoted to the derivation of linear conductance of the device in each of four regimes discussed earlier. A brief discussion on the thermodynamic observables of the double-grain device is given in Sec.~\ref{thdd}. Finally, we conclude in Sec.~\ref{diss}. Minor mathematical details of our calculations are given in the Appendices.

\section{Model}\label{model}
\begin{figure}[t]
\begin{center}
\includegraphics[scale=0.45]{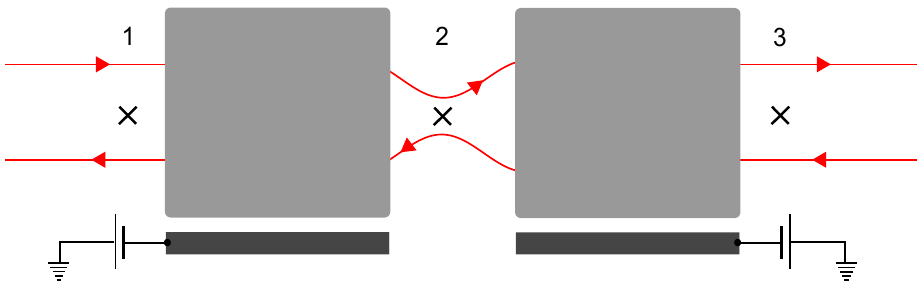}
\caption{Experimental implementation of the model Hamiltonian~\eqref{sg9}. Gate tunable double-charge grains are connected to each other and also to the metallic leads by fully tunable quantum point contacts (see text for details).}\label{fig1}
\end{center}
\end{figure}
The schematic of the experimental setup which consists of two hybrid metal-semiconductor islands is shown in Fig.~\ref{fig1}. Each island hosts a macroscopically-large number of charge states which can be tuned by applying voltage on the corresponding nearby gates. The islands are connected to each other and also to the metallic leads by fully tunable quantum point contacts (QPCs) represented by the $\times$ signs in the figure. Non-interacting electrons in these QPCs can be modeled as pairs of counter-propagating quantum Hall edges which are partially covered by two metallic islands.

The treatment of chiral edge states is greatly simplified by bosonizing fermion
operators. In the bosonic representation, the spin-polarized electrons in three interspaced
QPCs can be described by the quadratic Hamiltonian of the form~\cite{Wen_1990}
\begin{align}
H_0=\frac{v_{\rm F}}{4\pi}\sum_{\alpha=1}^3\int^\infty_{-\infty}\!\!\! dx\Big[\left(\partial_x\Phi_{\alpha, {\rm R}}\right)^2+\left(\partial_x\Phi_{\alpha, {\rm L}}\right)^2\Big],\label{sg4}
\end{align}
where $\alpha$ is the QPCs number index, $v_{\rm F}$ stands for the Fermi velocity~\footnote{We assume the same Fermi velocity $v_{\rm F}$ for each edge channel.}, and $\Phi_{\alpha, {\rm L/R}}(x, t)$ represents the bosonic field corresponding to the incoming/outgoing chiral fermions with charge density
\begin{align}\label{sg5}
\rho_{\alpha, {\rm L/R}}(x, t)=\frac{1}{2\pi}\partial_x\Phi_{\alpha, {\rm L/R}}(x, t).
\end{align}

As in the experiment~\cite{ccd}, we consider two identical grains each having charging energy $E_c$. As discussed earlier in the introductory section, since the two grains in Fig.~\ref{fig1} are connected by a short electronic channel implemented by the QPC, both intra and inter-grain Coulomb interactions play important roles. Therefore, we describe the charging effects on the device using the constant interaction model Hamiltonian
\begin{align}\label{sg6}
H_c=E_c\left[\left(\hat{\widetilde{\mathit{N}}_{1}}\right)^2+\left(\hat{\widetilde{\mathit{N}}_{2}}\right)^2\right]+2 E_I \hat{N}_1\hat{N}_2,
\end{align}
where $\hat{\widetilde{\mathit{N}}_{i}}=\hat{N}_i-\ngg$, and we assumed the same gate voltage $\ngg$ for both grains. In Eq.~\eqref{sg6}, both interactions energies $E_c$ and $E_I$ are assumed to be larger than the mean level spacing of the grains, which we set to be zero in the rest of the discussions. The number density operators $\hat{N}_{1, 2}$ of the two grains can be expressed in terms of charge densities as~\cite{Slobodeniuk_2013}
\begin{align}\label{sg7}
\hat{N}_i=\int^\infty_0 \!\!dx\;\delta\rho_i(x, t)+\int^0_{-\infty} \!\!dx\;\delta\rho_{i+1}(x, t),
\end{align}
where we introduced the density difference operator $\delta\rho_i=\rho_{i,{\rm R}}-\rho_{i,{\rm L}}$.

When all QPCs are tuned to the perfect transmission, the Hamiltonians $H_0$, and $H_c$ formally complete the construction of a model Hamiltonian of the device. However, as discussed earlier, in this limit the CB oscillations are completely washed out as the flow of electrons becomes continuous and there is no charge granularity in the system. Therefore, we introduce weak backscattering in each QPC described by the non-linear Hamiltonian
\begin{equation}\label{sg8}
H_B=\sum_{\alpha=1}^3\frac{D|r_\alpha|}{\pi}\cos\left[\Phi_{\alpha, {\rm R}}(0, t)-\Phi_{\alpha, {\rm L}}(0, t)\right],
\end{equation}
where $|r_\alpha|$ stands for the bare reflection amplitude of the QPC $\alpha$, and $D$ is the high energy cutoff introduced in the bosonization framework. We note that the interaction in our problem is only the charging energy, and thus the bandwidth $D$ should not appear in any physical quantities. Finally, we separate the total Hamiltonian into two parts
\begin{equation}\label{sg9}
H=H'+H_B,
\end{equation}
where $H'$ is the quadratic Hamiltonian defined by $H'=H_0+H_c$. To make further progress, we will treat $H'$ exactly while the non-linear Hamiltonian $H_B$ will be accounted for by perturbation theory to leading order in backscattering amplitudes $|r_\alpha|$.

\section{Scattering theory of boson}\label{sth}
In a fully ballistic setting, as discussed earlier, the flow of electrons becomes continuous. The Hamiltonian $H'$ then represents the plasmonic excitations traveling along the chiral edges. To describe the scattering of plasmonic excitations, we follow the equation of motion approach presented in great detail in Refs.~\cite{Slobodeniuk_2013,Sukhorukov_2016,morel2021, KBM}. As an illustration of the method in the context of our problem, we first write the Heisenberg equation of motion for the bosonic field $\Phi_{\alpha, \sigma}$, where $\sigma={\rm L/R}=\mp$, in the form $\partial_t\Phi_{\alpha, \sigma}(x, t)=i\left[H'(t), \Phi_{\alpha, \sigma}(x, t)\right]$. This equation can be simplified by using Eqs.~\eqref{sg4}$-$\eqref{sg7} followed by the application of standard identity
\begin{align}\label{sg10}
\left[\partial_x\Phi_{\alpha, \sigma}(x, t), \Phi_{\beta, \sigma'}(y, t)\right]=2\pi i\sigma\delta_{\alpha\beta}\delta_{\sigma\sigma'}\delta(x-y).
\end{align}
The equations of motion for the bosonic fields corresponding to the first QPC are
\begin{align}\label{sg11}
\partial_t\Phi_{1, {\rm L/R}}(x, t)&=\pm v_{\rm F}\partial_x\Phi_{1, {\rm L/R}}(x, t)\nonumber\\
&\quad\;-2\left[E_c \hat{\widetilde{\mathit{N}}_{1}}(t)+E_I\hat{N}_2(t)\right]\theta(x),
\end{align}
and similarly for other QPCs. From Eq.~\eqref{sg11}, it is seen that the Coulomb interaction between the grains mediates the coupling of bosonic fields on the leftmost part of the device to the charge state of the distant grain, the second island. This effect caused by Coulomb coupling will have crucial consequences on the nature of CB oscillations.

By solving Eq.~\eqref{sg11} for the bosonic field $\Phi_{1, {\rm L/R}}(0, t)$, and Fourier transforming the resulting expression into frequency space, we obtain the expression for $\Phi_{1, {\rm L/R}}(\omega)$ in terms of the incoming field $\Phi^{(0)}_{1, {\rm L/R}}(\omega)$ and number operators $\hat{N}_i(\omega)$. The number operators in the frequency space can be obtained from the equations~\eqref{sg5} and~\eqref{sg7} in the form
\begin{align}
\hat{N}_i(\omega)= &\frac{1}{2\pi}\Big[-\Phi_{i, {\rm R}}(\omega)+\Phi_{i, {\rm L}}(\omega)
\nonumber\\
&\qquad+\Phi_{i+1, {\rm R}}(\omega)-\Phi_{i+1, {\rm L}}(\omega)\Big].\label{sg13}
\end{align}
Repeating these steps also for the bosonic fields of the remaining two QPCs, $\Phi_{2, {\rm R/L}}$ and $\Phi_{3, {\rm R/L}}$, followed by the application of Eq.~\eqref{sg13}, we eventually get six linear equations for the bosonic fields in the frequency representation. Those six equations can be solved to express each $\Phi_{\alpha, {\rm R/L}}(\omega)$ as the linear combination of six incoming fields $\Phi^{(0)}_{\alpha, {\rm R/L}}(\omega)$ in the form~\cite{KBM}
\begin{align}\label{sg14}
\Delta\Phi(\omega)=\mathbb{M}(\omega)\Delta\Phi^{(0)}(\omega)+\mathbb{N}(\omega),
\end{align}
where the matrix $\mathbb{M}$ is given in the appendix~\ref{scmm}, and the matrix of the bosonic fields is defined as
\begin{align}
\Delta\Phi &=\left(\Phi _{\rm 1, R}, \Phi _{\rm 1, L}, \Phi _{\rm 2, R}, \Phi _{\rm 2, L}, \Phi _{\rm 3, R}, \Phi _{\rm 3, L}\right)^{\mathbb{T}},\label{sg15}
\end{align}
with $\Delta\Phi^{(0)}=\Delta\Phi(\Phi_{\alpha, {\rm R/L}}\to \Phi^{(0)}_{\alpha, {\rm R/L}})$. In Eq.~\eqref{sg14}, we also introduced the matrix $\mathbb{N}$ containing the information of the gate voltage
\begin{align}
\mathbb{N}(\omega)&=4\pi^2 \delta(\omega)\widetilde{\mathit{N}}_{\rm g}\left(\begin{array}{cccccc}
 0, 1, 1, 1,  1, 0 \\
\end{array}
\right)^{\mathbb{T}},\label{sg17}
\end{align}
where the renormalized gate voltage due to the Coulomb coupling between the grains is given by
\begin{align}
\widetilde{\mathit{N}}_{\rm g}=\left(\frac{E_c}{E_c+E_I}\right)\ngg,\label{sg17a}
\end{align}
and such a renormalization will affect the periodicity of CB oscillations.

The main advantage of Eq.~\eqref{sg14} is that all bosonic correlation functions and commutators required for the calculation of transport properties can easily be obtained from those of corresponding incoming fields $\Phi^{(0)} _{\alpha,{\rm , L/R}}$. For instance, the latter satisfies the usual commutation relation $\left[\Phi^{(0)}_{\alpha, {\rm R/L}}(t), \Phi^{(0)}_{\beta, {\rm R/L}}(t')\right]=-i\pi\delta_{\alpha\beta}\;{\rm sign}(t-t')$, and the correlation function
\begin{align}
\left<\!\Phi^{(0)}_{\alpha, {\rm R/L}}(\omega)\Phi^{(0)}_{\beta, {\rm R/L}}(\omega')\!\right>_{\!\!0}\!{=}\frac{4\pi^2\delta_{\alpha\beta}}{\omega'}n\!\left(\!\!\frac{\omega'}{T}\!\right)\!\!\delta\!\left(\omega+\omega'\right)\!,\label{sg18}
\end{align}
where $n(x)=\left(e^x-1\right)^{-1}$ is Bose distribution function. The equation~\eqref{sg14} essentially completes the construction of scattering theory of bosons applicable to the coupled hybrid metal-semiconductor islands.

\section{Charging energy modes}\label{chm}
To express Eq.~\eqref{sg14} in a more appealing form, we introduce a new bosonic field $\Phi_\alpha$ defined by
\begin{align}\label{sg19}
\Phi_\alpha(\omega)=\Phi_{\rm \alpha, R}(\omega)-\Phi_{\rm \alpha, L}(\omega).
\end{align}
From equations.~\eqref{sg14} and~\eqref{sg19}, we get the equation of motion for the new bosonic field $\Phi_\alpha(\omega)$ in the form
\begin{align}
\left[\!\!
\begin{array}{c}
\Phi _{1}(\omega) \\
 \Phi _{2}(\omega) \\
\Phi _{3}(\omega) \\
\end{array}\!\!
\right] &=\Bigg[\frac{\mathbb{J}_3}{3}+\mathbb{S}_a(\omega) +\mathbb{S}_b(\omega)\Bigg]\left[\!\!
\begin{array}{c}
 \Phi^{(0)} _{1}(\omega) \\
 \Phi^{(0)} _{2}(\omega) \\
 \Phi^{(0)} _{3}(\omega) \\
\end{array}\!\!
\right]\nonumber\\
&\qquad+4\pi^2 \widetilde{\mathit{N}}_{\rm g}\delta(\omega)\left[\!\!
\begin{array}{c}
 -1 \\
\phantom{-}0 \\
\phantom{-}1 \\
\end{array}
\right]\!\!,\label{sg20}
\end{align}
where $\mathbb{J}_3$ is the unit matrix with all entries being unity, and matrices $\mathbb{S}_{a, b}$ are defined as
\begin{align}
\mathbb{S}_a &=\mathcal{K}_a\left[
\begin{array}{ccc}
 \phantom{-}1 & \phantom{-}0 & -1 \\
\phantom{-} 0 & \phantom{-}0 & \phantom{-}0 \\
 -1 &\phantom{-}0 & \phantom{-}1 \\
\end{array}
\right],\label{sg21}\\
\mathbb{S}_b &=\mathcal{K}_b\left[
\begin{array}{ccc}
 \phantom{-}1 & -2  & \phantom{-}1 \\
 -2  & \phantom{-}4  & -2  \\
 \phantom{-}1 & -2  & \phantom{-}1 \\
\end{array}
\right].\label{sg22}
\end{align}
The functions $\mathcal{K}_{a, b}$ in Eqs.~\eqref{sg21} and~\eqref{sg22} directly encode the information about charging energy modes, and are defined by
\begin{align}
\mathcal{K}_a(\omega) &=\frac{1}{2i}\frac{\omega}{a-i\omega},\;\;a=\frac{E_+}{\pi},\;\;E_+=E_c+E_I,\label{sg23}\\
\mathcal{K}_b(\omega) &=\frac{1}{6i}\frac{\omega}{b-i\omega},\;\;b=\frac{3E_-}{\pi},\;\;E_-=E_c-E_I.\label{sg24}
\end{align}

From the equations~\eqref{sg23} and~\eqref{sg24}, we recognize the aforementioned charging modes as $(a, b)$, where the lighter/heavier mode is decided by the interaction parameters $E_c$ and $E_I$, and the mode $b$ vanishes at the typical value of Coulomb coupling $E_I=E_c$. Although this special point might be of only theoretical interest, we include this point as well in our theory. We note that our complete theory is not sensitive to the strengths of these modes. In addition, to obtain a close form expression of the linear conductance in the quantum and thermal regimes, it would be of no importance to have a precise identification of light and heavy charging energy modes. For the discussions relevant to the quantum-thermal mixed regime, however, we will be identifying those modes in a precise sense. 

The relation between the outgoing and incoming bosonic fields expressed by Eq.~\eqref{sg20} is one of the central equations which will be of use in large part of the following discussions. For the latter convenience, we thus present some of the implications of Eq.~\eqref{sg20}.

\subsection*{Implications of scattering solution}
Equation~\eqref{sg20} infers a remarkable decoupling of charging modes. The latter will result in the tremendous simplification in evaluating the bosonic correlators required to calculate the charge current, and to analyze different CB regimes discussed earlier.

When the strength of both charging modes is finite, i.e., $a\neq 0$, $b\neq 0$, the elements $\mathcal{K}_{a, b}$ in the limit of $\omega/a\to 0$, $\omega/b\to 0$ vanishes, and consequently $\mathbb{S}_{a, b}$ becomes null matrix. Then, in the absence of any external voltages, all outgoing bosonic fields become identical $\Phi_\alpha(\omega)=\sum_{\alpha=1}^3\Phi^{(0)}_{\alpha}(\omega)/3$. The latter corresponds to the situation where three quantum resistances are connected in series, i.e., conductance of the system measured at either left or right QPC in the absence of backscattering acquires the value $G_0=G_q/3$, where $G_q=1/2\pi$ is the quantum conductance.

Although the point $E_c=E_I$ might not be achievable in a real experiment, as a matter of principle it is worth considering this special point. In this case, for $\omega/a\to 0$ and in the absence of all external voltages, Eq.~\eqref{sg20} predicts
\begin{align}
\left[
\begin{array}{c}
\Phi _{1} \\
 \Phi _{2} \\
\Phi _{3} \\
\end{array}
\right]\to \frac{1}{2}\left[
\begin{array}{ccc}
1 &0&1 \\
 0&2&0 \\
1&0&1 \\
\end{array}
\right]\left[
\begin{array}{c}
\Phi^{(0)} _{1} \\
 \Phi^{(0)} _{2} \\
\Phi^{(0)} _{3} \\
\end{array}
\right].\nonumber
\end{align}
This equation implies that at the point $b=0$, the middle QPC completely decouples from the system, and the first and third QPCs from an equivalent circuit (this point will be further clarified in the next section). Consequently, the conductance in the absence of backscattering takes the value $G_0=G_q/2$.

From Eq.~\eqref{sg20}, it is seen that the bosonic field $\Phi_2$ corresponding to the middle QPC involves only the charging energy mode $b$, while that for side QPCs includes both modes. This is associated with the fact that the increased Coulomb coupling between the grains has the strongest influence on the connecting QPC. This whole point will be crucial for the discussions in the next section.

To arrive at Eq.~\eqref{sg20}, we assumed the absence of external voltage bias. The latter can be accounted for by manually adding voltage bias to the equation of motion. As our main focus is to calculate the linear conductance, the infinitesimal voltage $V$ can be accounted for by implementing gauge shifting of bosonic variables. Considering the above discussions, such a shift takes the form $\Phi_{\alpha}(t)\to\Phi_\alpha(t)-Vt/3$ for $b\neq 0$, and $\Phi_{1, 3}(t)\to\Phi_{1, 3}(t)-Vt/2$ for $b=0$, where $\Phi_2$ will be irrelevant at lest within the perturbative calculation of transport properties.

To evaluate charge current perturbatively in small backscattering amplitudes, which will be the subject of the next section, one would require the expression for the correlators $\mathcal{C}_{\alpha\beta}(t)=\left<e^{-i\Phi_\alpha(t)}e^{i\Phi_\beta(0)}\right>_{\!0}$. The equation~\eqref{sg20} then shows that only the off-diagonal correlators acquire gate voltage dependences, and hence CB oscillations in corresponding linear conductances.

Now we are in the position of discussing the evaluation of charge current by treating the non-linear Hamiltonian $H_B$ according to perturbation theory to leading order in backscattering amplitudes $|r_\alpha|$.

\section{Charge current}\label{chargec}
The current operator at the right output of the device at position $x=l$ can be defined as
\begin{equation}\label{sg26}
\hat{I}=-\frac{1}{2\pi}\partial_t\Phi_{3}(l, t).
\end{equation}
Application of infinitesimal voltage $V$ at the leftmost channel can be accounted for by shifting the bosonic field $\Phi_3(t)\to\Phi_3(t)-\left(\frac{1}{3}+\frac{\delta_b}{6}\right)V t$ (see previous section). Therefore, for $|r_\alpha|=0$, the linear conductance is given by
\begin{align}\label{sg27}
G_0=\frac{1}{2\pi}\left(\frac{1}{3}+\frac{\delta_b}{6}\right).
\end{align}

First non-vanishing correction to the average current due to finite backscattering $|r_\alpha|\neq 0$ is given by~\cite{KBM}
\begin{align}
\delta I={-}\int^t_{-\infty}\!\!\!\!\!\!dt_1\!\!\int^{t_1}_{-\infty}\!\!\!\! \!dt_2\!\;\Big<\! \Big[\left[\hat{I}(l, t), H_B(t_1)\right]\!\!, H_B(t_2)\Big]\!\Big>_{\!\!0}.\label{sg28}
\end{align}
For the evaluation of commutators in Eq.~\eqref{sg28}, first we re-express the backscattering Hamiltonian~\eqref{sg8} in slightly different form as
\begin{align}
H_B(t)=\sum_{\alpha=1}^3\Big[\mathcal{U}_\alpha(t)+\mathcal{U}^\dagger_\alpha(t)\Big],\;\mathcal{U}_\alpha\equiv\frac{D|r_\alpha|}{2\pi}e^{i\Phi_\alpha},\label{sg29}
\end{align}
and define the sum $\mathcal{U}(t)=\sum_{\alpha=1}^3\mathcal{U}_\alpha(t)$. Then, using the scattering matrix $\mathbb{M}$ given in the appendix and properly accounting for the $l$ dependence of the bosonic field $\Phi_{\alpha, {\rm L/R}}(l, t)=\Phi_{\alpha, {\rm L/R}}( t\pm t_0)$, where $t_0=l/v_{\rm F}\equiv l$, we write Eq.~\eqref{sg28} in the form
\begin{align}
\delta I=-\sum_{\alpha=1}^3\int^t_{-\infty}dt_1&\int^{t_1}_{-\infty} dt_2\; F_\alpha(t-t_1)\times\nonumber\\
&\Big[P_\alpha(t_1-t_2)+{\rm c.c.}\Big],\label{sg30}
\end{align}
where the function $	P_\alpha$ is defined by
\begin{align}
P_\alpha(t_1-t_2)=\left<\left[\mathcal{U}_\alpha(t_1), \mathcal{U}^\dagger(t_2)\right]\right>_0.\label{sg31}
\end{align}
In equation~\eqref{sg30} we also introduced another function $F_\alpha$ that depends on the position where the current is being measured, and is given as
\begin{align}
F_\alpha(t-t_1)=-\frac{1}{2\pi}\int^\infty_{-\infty} \!\!\!\!d\omega\Big[e^{i\omega t_0}\;\mathcal{V}_\alpha(\omega)+{\rm c.c.}\Big] e^{-i\omega(t-t_1)},\label{sg32}
\end{align}
The frequency dependent coefficients $\mathcal{V}_\alpha(\omega)$ is Eq.~\eqref{sg32} take the form
\begin{align}
\mathcal{V}_1 &=\frac{1}{3}-\mathcal{K}_a+\mathcal{K}_b,\nonumber\\
\mathcal{V}_2 &=\frac{1}{3}-2\mathcal{K}_b,\nonumber\\
\mathcal{V}_3 &=\frac{1}{3}+\mathcal{K}_a+\mathcal{K}_b\nonumber,
\end{align}
where $\mathcal{K}_{a, b}$ are defined in equations.~\eqref{sg23} and~\eqref{sg24}. We note that the coefficients $\mathcal{V}_\alpha$ satisfy the conservation law $\sum_\alpha\mathcal{V}_\alpha=1$, and the property $\mathcal{V}^*_\alpha(\omega)=\mathcal{V}_\alpha(-\omega)$. Besides, for $b=0$, the coefficient $\mathcal{V}_2$ vanishes. Therefore, by noticing this fact and closely examining equations~\eqref{sg20} and~\eqref{sg21}, one immediately concludes that the middle QPC is completely irrelevant for the special setting of $b=0$ as announced earlier.

With these understandings, it is straightforward to simplify Eq.~\eqref{sg30} to the form
\begin{align}
\delta I= &{-}\left(\!\frac{1}{3}{+}\frac{\delta_b}{6}\!\right)\!\!\int^\infty_{-\infty} \!\!\!\!\!dt \;e^{iV t\left(\frac{1}{3}{+}\frac{\delta_b}{6}\right)}\!\!\left<\left[\mathcal{T}^\dagger(t), \mathcal{T}(0)\right]\right>_0,\label{sg36}
\end{align}
where we introduced a new operator $\mathcal{T}(t)$ defined by
\begin{align}
\mathcal{T}(t)=\Big[\mathcal{U}_1(t)+\left(1-\delta_b\right)\mathcal{U}_2(t)+\mathcal{U}_3(t)\Big]_{V=0}.\label{sg37}
\end{align}
The prefactor in Eq.~\eqref{sg36} comes from the fact that $\mathcal{V}_{1, 3}(0)=(1/3+\delta_b/6)$ and $\mathcal{V}_{2}(0)=(1/3-\delta_b/3)$.

In the following discussions, we will be using Eqs.~\eqref{sg20} and~\eqref{sg36} for the evaluation of linear conductance. To accomplish the latter, we calculate the expressions for the bosonic correlators $\mathcal{C}_{\alpha\beta}(t)=\left<e^{-i\Phi_\alpha(t)}e^{i\Phi_\beta(0)}\right>_{\!0}$, substitute them into Eq.~\eqref{sg36}, take the voltage derivative of resulting expression in the limit of $V\to 0$, and finally perform the time integral. Using Eq.~\eqref{sg20} in combination with Eq.~\eqref{sg18}, one can easily evaluate the expressions for $\mathcal{C}_{\alpha\beta}(t)$, which takes the form
\begin{align}
\mathcal{C}_{\alpha\beta}(t)=e^{-2\pi i \widetilde{\mathit{N}}_{\rm g}\left(\alpha-\beta\right)}\exp\left[\mathcal{J}_{\alpha\beta}(t)\right],\label{sg38}
\end{align}
for the function $\mathcal{J}_{\alpha\beta}(t)$ given by
\begin{align}
\mathcal{J}_{\alpha\beta}(t)&=2\int^\infty_{-\infty}\frac{d\omega}{\omega}\frac{1}{1-e^{-\omega/T}}\Big[\left(e^{-i\omega t}-1\right)\mathcal{D}_{\alpha\beta}(\omega)\nonumber\\
&\qquad\qquad\qquad\qquad\;-\left(1-\delta_{\alpha\beta}\right)\mathcal{L}_{\alpha\beta}(\omega)\Big].\label{sg39}
\end{align}
To arrive at Eq.~\eqref{sg39}, we introduced new functions $\mathcal{D}_{\alpha\beta}(\omega)=\mathcal{D}_{\beta\alpha}(\omega)$ defined as
\begin{align}
\mathcal{D}_{11}(\omega)&=\mathcal{D}_{33}(\omega)=1-Z_a(\omega)-Z_b(\omega),\label{sg40}\\
\;\;\mathcal{D}_{22}(\omega) &=1-4Z_b(\omega),\label{sg41}\\
\mathcal{D}_{12}(\omega) &=\mathcal{D}_{23}(\omega)=2Z_b(\omega),\label{sg42}\\
\;\;\mathcal{D}_{13}(\omega) &=Z_a(\omega)-Z_b(\omega),\label{sg43}
\end{align}
with the function $Z_{a, b}$ given by
\begin{align}
Z_a(\omega) &=\frac{1}{2}-2\left|\mathcal{K}_a(\omega)\right|^2,\label{sg44}\\
Z_b(\omega) &=\frac{1}{6}-6\left|\mathcal{K}_b(\omega)\right|^2.\label{sg45}
\end{align}
Functions $\mathcal{L}_{\alpha\beta}(\omega)=\mathcal{L}_{\beta\alpha}(\omega)$ in Eq.~\eqref{sg39} take the form
\begin{align}
\mathcal{L}_{12}(\omega)&=\mathcal{L}_{23}(\omega)=1-\frac{Z_a(\omega)}{2}-\frac{9Z_b(\omega)}{2},\label{sg46}\\
\;\;\mathcal{L}_{13}(\omega)&=1-2Z_a(\omega).\label{sg47}
\end{align}
Having collected the complete information of correlators~\eqref{sg38}, it is now a straightforward task to calculate linear conductance from Eq.~\eqref{sg36}. The integrals~\eqref{sg39} can be calculated numerically to get the complete picture of linear conductance without considering any inequalities between temperature and charging energy modes. The next section will be devoted to the exploration of linear conductance in the different regimes discussed earlier.
\section{Linear conductance}\label{lc}
The bosonic correlator $\mathcal{C}_{\alpha\beta}(t)$ given by Eq.~\eqref{sg38} is fully responsible for the behavior of linear conductance. It contains information about two charging energy modes $(a, b)$ and temperature $T$. In addition, the charging energy modes in $\mathcal{C}_{\alpha\beta}(t)$ are decoupled. Therefore, in general, four different regimes can be defined as: i) quantum regime; $T\ll\left(a, b\right)$, ii) thermal regime; $T\gg\left(a, b\right)$, iii) quantum-thermal mixed regime; $b\ll T\ll a$ (this inequality will be discussed in more detail in the subsection~\ref{qtm}), and iv) the special point $b=0$.

As discussed earlier, when the charging energy mode $b$ vanishes, phase transition driven by Coulomb coupling takes place. In this case, the middle QPC which mediates the charge fluctuations between the two grains becomes completely irrelevant. The latter can most conveniently be seen by re-expressing the charging energy Hamiltonian~\eqref{sg6} to the form
\begin{align}
H_c=\frac{E_c{+}E_I}{2}\left(\hat{N}_1{+}\hat{N}_2\right)^2+\frac{E_c{-}E_I}{2}\left(\hat{N}_1{-}\hat{N}_2\right)^2,\label{sg48}
\end{align}
where we assumed the absence of gate voltage. Thus, for $E_c=E_I$, the charge fluctuations between two grains represented by the second term of Eq.~\eqref{sg48} vanishes, while the first term is identical to that of a single grain setup. We note that, in Eq.~\eqref{sg6} we assumed $E_c^{(1)}=E_c^{(1)}=E_c$, and thus light mode vanishes only at $E_c=E_I$. However, this is not only the possibility; by using charge grains of unequal size and hence the different charging energies, such a point can be engineered differently, as discussed in the introductory section. The CB oscillations in the linear conductance of a single grain device have already been explored in great detail in previous works~\cite{Matveev1995, Furusaki1995b, td}, and thus we do not discuss them here.

In the following, we explore CB oscillations in the linear conductance of the device in quantum regime, thermal regime, and quantum-thermal mixed regime separately~\footnote{Linear conductance in yet another regime, where temperature and the strengths of charging modes are identical, of the considered device can be evaluated analytically.}.

\subsection{Quantum regime} 
In the quantum regime, both charging energy modes are much larger than temperature; $T\ll\left(a, b\right)$. In this case, at the limit of $\omega/a\to 0$ and $\omega/b\to 0$, the matrices $\mathbb{S}_{a, b}(\omega)$ in Eq.~\eqref{sg20} vanish. Therefore, for $T\ll\left(a, b\right)$, the matrices $\mathbb{S}_{a, b}(\omega)$ can be integrated out. To illustrate the latter, we write the operator $\mathcal{U}_\alpha(t)$ given in Eq.~\eqref{sg29} in the form
\begin{equation}
\mathcal{U}_\alpha(t)=\frac{D|r_\alpha|}{2\pi}\exp\Big[{-}\frac{1}{2}\left<\Phi_\alpha^2(t)\right>_{{\!\rm HE}}\Big]e^{i\tilde{\Phi}_\alpha(t)},\label{sg49}
\end{equation}
where the free bosonic field $\tilde{\Phi}_\alpha$ is given by Eq.~\eqref{sg20} in the form
\begin{align}
\left[\!\!
\begin{array}{c}
\tilde{\Phi} _{1}(\omega) \\
 \tilde{\Phi} _{2}(\omega) \\
\tilde{\Phi} _{3}(\omega) \\
\end{array}\!\!
\right] &=\frac{\mathbb{J}_3}{3}\left[\!\!
\begin{array}{c}
 \Phi^{(0)} _{1}(\omega) \\
 \Phi^{(0)} _{2}(\omega) \\
 \Phi^{(0)} _{3}(\omega) \\
\end{array}\!\!
\right]+4\pi^2 \widetilde{\mathit{N}}_{\rm g}\delta(\omega)\left[\!\!
\begin{array}{c}
 -1 \\
\phantom{-}0 \\
\phantom{-}1 \\
\end{array}
\right]\!\!.\label{sg50}
\end{align}
From the equations~\eqref{sg18}, ~\eqref{sg21} and~\eqref{sg22}, the integration of high energy modes results in~\cite{KBM}
\begin{align}
\left<\Phi^2_1(t)\right>_{\rm HE} &=\left<\Phi^2_3(t)\right>_{\rm HE}=-\frac{4}{3}\ln\left(\frac{3^{1/4}\tilde{E} e^\gamma}{\pi D}\right),\label{sg51}\\
\left<\Phi^2_2(t)\right>_{\rm HE} &=-\frac{4}{3}\ln\left(\frac{3E_- e^\gamma}{\pi D}\right),\label{sg52}
\end{align}
where $\gamma$ stands for the Euler's constant, and we introduced a new symbol $\tilde{E}_c$ representing the effective charging energy
\begin{align}\label{sg53}
\tilde{E}_c=E_+^{3/4} E^{1/4}_-.
\end{align}

Therefore, in the view of the equations~\eqref{sg49}$-$\eqref{sg52}, integrating out the high energy charging modes is equivalent to the renormalization of QPCs reflection coefficients
\begin{align}
|r_{1, 3}|\to|r'_{1, 3}| &=|r_{1, 3}|\left[\frac{3^{1/4} e^\gamma \tilde{ E}_c}{\pi D}\right]^{2/3},\label{sg54}\\
|r_{2}|\to|r'_{2}| &= |r_2|\sqrt{\frac{E_-}{E_+}}\left[\frac{3e^\gamma \tilde{E}_c }{\pi D}\right]^{2/3},\label{sg55}
\end{align}
and replacing
\begin{align}
\mathcal{D}_{\alpha\beta}(\omega) &\to \mathcal{D}_{\alpha\beta}(0)=\frac{1}{3},\nonumber\\
\mathcal{L}_{\alpha\beta}(\omega) &\to \mathcal{L}_{\alpha\beta}(0)=0,\nonumber
\end{align}
in the expression of the integral $\mathcal{J}_{\alpha\beta}(t)$ given in Eq.~\eqref{sg39}. The resulting integral $\mathcal{J}_{\alpha\beta}(t)$ represents the free bosonic correlator, which is given by
\begin{align}
\exp\left[\mathcal{J}_{\alpha\beta}(t)\right]&=e^{-i\pi/3}\left[\frac{\pi T}{D \sinh\pi T\left(t-i\delta^+\right)}\right]^{2/3}.\label{sg58}
\end{align}

By combining Eq.~\eqref{sg58} with Eq.~\eqref{sg38}, we access the correlation functions $\mathcal{C}_{\alpha\beta}(t)$. Finally, we substitute $\mathcal{C}_{\alpha\beta}(t)$ into Eq.~\eqref{sg36}, take the voltage differentiation in the limit $V\to 0$, and perform the elementary time integrations to get an expression for the correction to linear conductance $\delta G$ due to finite backscattering amplitudes $|r_\alpha|\neq 0$. Adding the correction $\delta G$ with unperturbed conductance given by Eq.~\eqref{sg27}, we arrive at the final expression of linear conductance in the quantum regime
\begin{align}
G(T)=\frac{1}{6\pi}\!\!\left[\!1-\frac{3^{1/3}\!\sqrt\pi }{6}\frac{\Gamma[1/3]}{\Gamma[5/6]}\Lambda(\widetilde{\mathit{N}}_{\rm g})\!\!\left(\!\frac{e^\gamma \tilde{E}_c}{\pi^2 T}\!\right)^{\!4/3}\right]\!\!,\label{sg59}
\end{align}
where the function $\Lambda(\widetilde{\mathit{N}}_{\rm g})$ is determined by the gate voltage and the reflection amplitudes of the QPCs, and is given by
\begin{align}
\Lambda(\widetilde{\mathit{N}}_{\rm g}) &=|r_1|^2+3 |\tilde{r}_2|^2+|r_3|^2+2 |r_1||r_3| \cos (4 \pi \widetilde{\mathit{N}}_{\rm g})\nonumber\\
&\;\;\;\;+2 \sqrt{3} |\tilde{r}_2| (|r_1|+|r_3|) \cos (2 \pi  \widetilde{\mathit{N}}_{\rm g}),\label{sg60}
\end{align}
with renormalized reflection amplitude of connecting QPC
\begin{align}
|\tilde{r}_2|=|r_2|\sqrt{\frac{E_-}{E_+}}.\label{sg61}
\end{align}
The expression of conductance~\eqref{sg59} retrieves the previous result in the limit of $E_+=E_-$, i.e., in the absence of inter-grain interaction $E_I=0$~\cite{KBM}. Equation~\eqref{sg59} shows that the Coulomb coupling between the grains introduces an effective energy scale given by Eq.~\eqref{sg53}, and also renormalizes the reflection coefficient of connecting QPC expressed by Eq.~\eqref{sg61}. In addition, the presence of $E_I$ also affects the peridicity of CB oscillations as can be seen from the equations~\eqref{sg59} and~\eqref{sg17a}.
\subsection{Thermal regime}
In the thermal regime, the temperature is much greater than both of the charging energy modes, i.e., $T\gg\left(a, b\right)$. To calculate the linear conductance in this regime, we first discuss the evaluation of diagonal and off-diagonal correlators~\eqref{sg38} separately.

The diagonal correlators $\mathcal{C}_{\alpha\alpha}(t)$, by recognizing the identity
\begin{align}
\exp\!\Big[2\!\int^\infty_{-\infty}\!\! \frac{d\omega}{\omega}\frac{e^{-i\omega t}{-}1}{1{-}e^{-\omega/T}}\Big]\!={-}\!\left(\!\frac{\pi T}{D}\right)^2\!\!\frac{1}{\sinh^2\!\pi T(t{-}i\delta^+)},\nonumber
\end{align}
can be expressed as
\begin{align}
\mathcal{C}_{\alpha\alpha}(t)=e^{\mathcal{J}_{\alpha\alpha}(t)}&=-\!\left(\!\frac{\pi T}{D}\right)^2\!\!\frac{\tilde{\mathcal{D}}_{\alpha\alpha}(t)}{\sinh^2\!\pi T(t{-}i\delta^+)},\label{sg62}
\end{align}
with the diagonal function $\tilde{\mathcal{D}}_{\alpha\alpha}(t)$ defined by 
\begin{align}
\tilde{\mathcal{D}}_{\alpha\alpha}(t)=\exp\Big[2\int^\infty_{-\infty}\!\! \frac{d\omega}{\omega}\frac{e^{-i\omega t}-1}{1-e^{-\omega/T}}\left[\mathcal{D}_{\alpha\alpha}(\omega){-}1\right]\Big].\label{sg63}
\end{align}
Then we evaluate the integral $\tilde{\mathcal{D}}_{\alpha\alpha}(t)$ by expanding the Bose function in small $\omega/T$. As an example, for $\tilde{\mathcal{D}}_{11}(t)$ we get~\cite{sochin}
\begin{align}
\tilde{D}_{11}(t) &\simeq \exp\left[\frac{1}{2}(E_++E_-)(T t^2+i t)\right],\label{sg64}
\end{align}
and similar results for all diagonal functions $\tilde{D}_{\alpha\alpha}(t)$. The equations~\eqref{sg62} and~\eqref{sg38} then give the desired result for the diagonal parts of the conductance correction corresponding to the three QPCs (see Ref.~\cite{td} for the details)
\begin{align}
\delta G_{\alpha\alpha}=-\frac{|r_\alpha|^2}{18\pi}.\label{sg65}
\end{align}

Now we turn our attention to the evaluation of off-diagonal correlators~\eqref{sg38}. To accomplish the latter, we note that the integrals $\mathcal{J}_{\alpha\beta}$ for $\alpha\neq \beta$ can be separated into time-depended and time-independent parts. As an illustration, we write $\mathcal{J}_{13}$ as
\begin{align}
\exp\left[\mathcal{J}_{13}(t)\right] &=\exp\Bigg[2\int^\infty_{-\infty}\!\! \frac{d\omega}{\omega}\frac{e^{-i\omega t}{-}1}{1{-}e^{-\omega/T}}\left[Z_a(\omega){-}Z_b(\omega)\right]\nonumber\\
&\qquad\;\;\;\;-2\!\int^\infty_{-\infty}\!\!\! \frac{d\omega}{\omega}\frac{1}{1{-}e^{-\omega/T}}\frac{\omega^2}{\omega^2{+}a^2}\Bigg].\label{sg66}
\end{align}
The first part of Eq.~\eqref{sg66} gives an exponential factor, as in Eq.~\eqref{sg64}, and the second part is a standard integral (which needs to be regularized, and thus depends on the bandwidth). In the thermal regime, Eq.~\eqref{sg66} gives
\begin{align}
\exp\left[\mathcal{J}_{13}(t)\right]&=\left(\frac{2\pi T}{D}\right)^2\exp\left(-\frac{2\pi^2 T}{E_+}\right)\times\nonumber\\
&\;\;\;\;\exp\left[-\frac{1}{2}(E_+{-}E_-)(T t^2+i t)\right].\label{sg67}
\end{align}
By using the equations~\eqref{sg36},~\eqref{sg38} and~\eqref{sg67}, we finally get the off-diagonal corrections to the linear conductance
\begin{align}
\delta G_{13}+\delta G_{31} &=\mathcal{A}\int^\infty_{-\infty} \!\!\!dt\;t\;\exp\left[{-}\frac{T t^2}{2}(E_+{-}E_-)\right]\times\nonumber\\
& \qquad\qquad\qquad\;\sin\!\left[\!\frac{(E_+{-}E_-)t}{2}\right],\label{sg68}
\end{align}
where the prefactor in the above integral is defined by
\begin{align}
\mathcal{A}=\frac{-4|r_1||r_3|}{9}\cos(4\pi\widetilde{\mathit{N}}_{\rm g})T^2\exp\left(\!\!{-}\frac{2\pi^2 T}{E_+}\right).\label{sg69}
\end{align}
Equation~\eqref{sg68} shows that the corrections to the cross-conductances in thermal regime, between the distant QPCs (QPC 1 and QPC 3), vanish in the absence of inter-grain Coulomb coupling, i.e., when $E_+=E_-$. This illustrates another importance of considering inter-grain interaction for the study of CB oscillations in a realistic double-grain device. Evaluation of the standard integral in Eq.~\eqref{sg68} results in, 
\begin{align}
\delta G_{13}+\delta G_{31}=-\frac{1}{18\pi}\;2|r_1||r_3|\cos\left(4\pi \widetilde{\mathit{N}}_{\rm g}\right)\mathcal{W}(T),\label{sg70}
\end{align}
where we introduce a new function $\mathcal{W}(T)$ defined by
\begin{align}
\mathcal{W}(T)=\left(1{-}\delta_{E_I}\right)2\pi\sqrt{\frac{\pi T}{E_I}}\exp\left(-\frac{2\pi^2 T}{E_+}\right).\label{sg71}
\end{align}

Repeating the same procedure as outlined above, one can calculate the corrections to the remaining cross-conductances. Collecting all contributions gives the final results for the correction to the linear conductance in the thermal regime 
\begin{align}
G(T)&=\frac{1}{6\pi}\Big[1-\frac{1}{3}\Big\{|r_1|^2+|r_2|^2+|r_3|^2\nonumber\\
&\qquad\;\;\;\;\;\;+2|r_2|\left(|r_1|{+}|r_3|\right)\cos(2\pi \widetilde{\mathit{N}}_{\rm g})\mathcal{X}(T)\nonumber\\
&\qquad\quad\;\;\;+2|r_1||r_3|\cos(4\pi \widetilde{\mathit{N}}_{\rm g})\mathcal{W}(T)\Big\}\Big],\label{sg72}
\end{align}
with the function $\mathcal{X}(T)$ given by
\begin{align}
\mathcal{X}(T)=2\pi\sqrt{\frac{\pi T}{E_-}}\exp\left[{-}\frac{\pi^2 T}{2}\left(\frac{1}{E_+}{+}\frac{1}{E_-}\right)\right].\label{sg73}
\end{align}
Conductance~\eqref{sg72} shows that the thermal decay of CB oscillations contributed by the function $\mathcal{W}(T)$ vanishes at the typical value of gate voltages $\widetilde{\mathit{N}}_{\rm g}=\left(4n+1\right)/8$, while that by the function $\mathcal{X}(T)$ vanishes at $\widetilde{\mathit{N}}_{\rm g}=\left(2n+1\right)/4$. In addition, $\mathcal{W}(T)$ remains finite only because of the Coulomb coupling between the grains. The latter is obvious already at the level of equation of motion Eq.~\eqref{sg11}, where we see that $E_I\neq 0$ mediates the coupling between the bosonic fields at the leftmost part of the device with that at the rightmost part.

\subsection{Quantum-thermal mixed regime}\label{qtm}
In the previous discussions of conductance in quantum and thermal regime, both charging modes are either much larger or much smaller than the base temperature of the device. Therefore, it was not necessary to identify which mode is lighter one. In this subsection, we are looking for the possibility that one mode is much smaller and the other is much larger than the temperature. Since the two modes $a$ and $b$ discussed in the section~\ref{chm} are not evenly distributed among all QPCs, it is crucial to precisely identify the light and heavy charging modes for the discussion of quantum-thermal mixed regime.

From equations~\eqref{sg23} and~\eqref{sg24}, we have the expressions of the charging modes as $a=(E_c+E_I)/\pi$ and $b=3(E_c-E_I)/\pi$, and they are identical for $E_I=E_c/2$. The mode $b$ decreases with increasing $E_I$ while the other mode increases for the same. At the limit of strong Coulomb coupling $E_I/E_c\sim 1$, the regime $b\ll a$ can be identified. Therefore, we define the quantum-thermal mixed regime such that $b\ll T\ll a$. Obviously, for two identical grains without Coulomb interactions between them, this regime does not exist, and hence the quantum-thermal regime of the device~\ref{fig1} is the sole consequence of Coulomb coupling between the grains. It is important to note that the quantum-thermal regime may also arises in the absence of Coulomb coupling if the charge grains are of reasonably different sizes as detailed in appendix~\ref{diffch}. In addition, this regime naturally arises in a device where the charge grains have considerably different charging energies and are connected through a tunnel barrier (see also Ref.~\cite{mkk}).

To study the linear conductance in the mixed regime $b\ll T\ll a$, we first performed the frequency integration in the limit of $T\ll a$ for the part of Eq.~\eqref{sg39} that only depends on $|\mathcal{K}_a|^2$. It is important to note from Eq.~\eqref{sg20} that the bosonic fields $\Phi_{1, 3}$ corresponding to the left and right QPCs contain both modes $a$ and $b$, while that of the middle QPC contains only the light mode $b$. Therefore, in this mixed regime, the middle QPC always remains in the thermal regime, and hence the frequency integration of $|\mathcal{K}_a|^2$ does not influence the bosonic field $\Phi_2$. However, such integration will renormalize the reflection amplitudes $|r_{1, 3}|$. After finding these renormalized reflection amplitudes, the rest of the calculations for the correction to linear conductance would be formally identical to that of the thermal regime presented in great detail in the previous subsection. Therefore, we just quote the final result for the linear conductance in quantum-thermal mixed regime
\begin{align}
G(T) &=\frac{1}{6\pi}\Bigg[1-\Pi_{\rm q}(\widetilde{\mathit{N}}_{\rm g})\left(\frac{e^\gamma E_+}{\pi^2 T}\right)-\Pi_{\rm t}(\widetilde{\mathit{N}}_{\rm g})\Bigg],\label{sg76}
\end{align}
where, $\Pi_{\rm q}$ encodes the purely quantum contribution defined by
\begin{align}
\Pi_{\rm q}(\widetilde{\mathit{N}}_{\rm g}) &=\frac{\pi}{6}\Big[|r_1|^2{+}|r_3|^2{+}2|r_1||r_3|\cos\left(4\pi \widetilde{\mathit{N}}_{\rm g}\right)\Big].\label{sg77}
\end{align}
The thermal oscillations in Eq.~\eqref{sg76} are expressed as
\begin{align}
\Pi_{\rm t}(\widetilde{\mathit{N}}_{\rm g})=\frac{|r_2|^2}{3}{+}\frac{|r_2|}{3}\left(|r_1|{+}|r_3|\right)\cos\left(\!2\pi \widetilde{\mathit{N}}_{\rm g}\right)\!\mathcal{Y}(T),\label{sg78}
\end{align}
for the temperature dependent function
\begin{align}
\mathcal{Y}(T)=2\pi\sqrt{\frac{2 e^\gamma E_+}{\pi E_-}}\exp\left(-\frac{\pi^2 T}{2E_-}\right).\label{sg79}
\end{align}

The behavior of linear conductance~\eqref{sg76} in the quantum-thermal mixed regome is markedly different as compared to any of the corresponding results obtained earlier in this work. The remarkable separation of purely quantum contribution from the thermal part results in a number of effects. For fully ballistic middle QPC $|r_2|=0$, the entire thermal contribution expressed in Eq.~\eqref{sg78} vanishes. In this case, the temperature correction to the linear conductance~\eqref{sg76} takes the similar form as that in the case of a single island setup. However, as indicated by Eq.~\eqref{sg59}, the scaling behavior of the linear conductance in the quantum regime is not affected by setting $|r_2|=0$. Therefore, our results show that Coulomb interaction dependent scaling behavior of linear conductance might be observed in the double-grain device considered in this work.

At the typical value of gate voltage $\widetilde{\mathit{N}}_{\rm g}^*=(2n+1)/4$ and symmetric left and right contacts $|r_1|=|r_3|$, entire CB oscillations in the linear conductance vanishes. In this case, conductance becomes constant $G=\left(1-|r_2|^2/3\right)/6\pi$, which is affected merely by the reflection amplitude of the middle QPC. The latter also holds for $|r_1|=|r_3|=0$ at any gate voltages. However, with the left and right contacts being asymmetric $|r_1|\neq |r_3|$, and the gate voltage is still tuned to $\widetilde{\mathit{N}}_{\rm g}^*$, only the thermal decay of CB oscillations given by the second term of Eq.~\eqref{sg78} vanishes. 

\section{Thermodynamics}\label{thdd}
In this section, we provide a brief discussion on the evaluation of thermodynamic properties of the setup~\ref{fig1}. The central ingredient to calculating thermodynamic observables is the system's free energy, which is defined by $\mathcal{E}=-T\ln\mathcal{Z}$, where $\mathcal{Z}$ is the partition function. $\mathcal{E}$ can easily be accessed in the same way as for the linear conductance discussed in previous sections. To get the first non-vanishing correction $\delta\mathcal{E}$ to the free energy due to small backscattering amplitudes $|r_\alpha|$, one requires the expression for the correlator $\left<H_B(t)H_B(0)\right>_{\!0}$. The latter is fully characterized by Eq.~\eqref{sg38}. Therefore, $\delta\mathcal{E}$ in all the regimes discussed earlier can be calculated straightforwardly. As an illustration, the free energy correction in the quantum regime $T\ll(a, b)$ takes the form
\begin{align}
\delta \mathcal{E}/T=-\frac{3^{5/6} }{4\sqrt\pi}\frac{\Gamma[1/3]}{\Gamma[5/6]}\Lambda(\widetilde{\mathit{N}}_{\rm g})\!\!\left(\!\frac{e^\gamma \tilde{E}_c}{\pi^2 T}\!\right)^{\!4/3}\!\!,\label{sg80}
\end{align}
and contains the same gate voltage dependence as in the corresponding linear conductance~\eqref{sg59}.

\section{Conclusion}\label{diss}
We developed a theoretical framework to evaluate the transport properties of a nanoelectronic device consisting of coupled hybrid metal-semiconductor islands. Our theory took into account both intra and inter-island Coulomb interactions exactly while treating backscattering in quantum point contacts perturbatively. In the strong tunneling regime of the double-island device, our theory works well without any restrictions on the relation between base temperature and the strength of Coulomb interactions. We highlighted a number of effects arising from the Coulomb coupling, and discussed their importance in the transport and thermodynamic properties of a realistic double-island device.

We showed that the linear conductance of the considered device depends heavily on the strengths of the two eigen-modes $(a, b)$ of the charging energy. Depending on the strength of these eigen-modes and the temperature, we identified three distinct operations regimes of the device; i) quantum regime $T\ll (a, b)$, ii) thermal regime $T\gg (a, b)$, and iii) quantum-thermal mixed regime $b\ll T\ll a$. The accidental vanishing of one of the charging modes, and hence a sharp phase transition, has also been accounted for within our theoretical framework. We applied our theory to evaluate the linear conductances in each operational regime of the device, and revealed their distinctive Coulomb blockade oscillations. While the emergence of the quantum-thermal mixed regime in the considered device is a sole consequence of the strong inter-island Coulomb interactions, we reasoned its relevance also in other mesoscopic devices containing double charge grains.

\section*{Acknowledgements}
Fruitful discussions with K. A. Matveev are gratefully acknowledged. This work was supported by the U.S. Department of Energy, Office of Science, Basic Energy Sciences, Materials Sciences and Engineering Division.

\appendix
\section{Expression for the scattering matrix}\label{scmm}
The matrix $\mathbb{M}$ in Eq.~\eqref{sg14} expressing the bosonic fields $\Phi_{\alpha, {\rm R/L}}(\omega)$ as the linear combination of six incoming fields $\Phi^{(0)}_{\alpha, {\rm R/L}}(\omega)$ takes the form
\begin{align}
\mathbb{M}=\left[\!
\begin{array}{cccccc}
 1 & 0 & 0 & 0 & 0 & 0 \\
 y{+}z & 1{-}y{-}z & -2 z & 2 z & z-y & y{-}z \\
 y{+}z & -y-z & 1{-}2 z & 2 z & z-y & y{-}z \\
 y{-}z & z-y & 2 z & 1{-}2 z & -y-z & y{+}z \\
 y{-}z & z-y & 2 z & -2 z & 1{-}y{-}z & y{+}z \\
 0 & 0 & 0 & 0 & 0 & 1 \\
\end{array}\!
\right],\nonumber
\end{align}
where 
\begin{align}
y(\omega)=\frac{1}{2}-\mathcal{K}_a(\omega),\;\;z(\omega)=\frac{1}{6}-\mathcal{K}_b(\omega),
\end{align}
with $\mathcal{K}_{a, b}(\omega)$ as defined in equations~\eqref{sg23} and~\eqref{sg24}.
\section{Scattering solution for $E^{(1)}_c\neq E^{(2)}_c$}\label{diffch}
We assume the situation when the charge grains are of considerably different sizes. For simplicity, we also set $E_I=0$, $\ngg=0$ and rename the charging energies as $E^{(1)}_c=e_1$ and $E^{(2)}_c=e_2$. Following the procedures outlined in Sec~\ref{sth}, we arrive at the similar equation as Eq.~\eqref{sg20} for the scattering solution with $\mathbb{S}_{a, b}$ replaced by $\mathbb{S}_{o, e}$, where
\begin{align}
\mathbb{S}_e=\!\frac{\tilde{\mathbb{S}}_e(e_1, e_2)}{3 i (\eta_e{-}\eta_o)}\frac{\omega }{\eta_e{-}i \omega },\;\mathbb{S}_o=\!\frac{\tilde{\mathbb{S}}_o(e_1, e_2)}{3 i (\eta_e{-}\eta_o)}\frac{\omega }{\eta_o{-}i \omega }.\label{pp1}
\end{align}
In the equation~\eqref{pp1}, the charging energy modes $\eta_{e/o}$ are defined by
\begin{align}
\eta_{e/o}=\frac{1}{\pi}\frac{e_1+e_2}{2}\left[2\pm\sqrt{1+3\left(\frac{e_1-e_2}{e_1+e_2}\right)^2}\right],\label{md1}
\end{align}
\begin{widetext}
\noindent and the matrix $\tilde{\mathbb{S}}_{e}$ takes the form
\begin{align}
\tilde{\mathbb{S}}_e =&\frac{1}{\pi}\left[\!
\begin{array}{ccc}
 -e_1-4 e_2 &\;\;\; 2 e_2-e_1 &\;\;\; 2 (e_1+e_2) \\
 2 e_2-e_1 &\;\;\; -e_1-e_2 &\;\;\; 2 e_1-e_2 \\
 2 (e_1+e_2) &\;\;\; 2 e_1-e_2 &\;\;\; -4 e_1-e_2 \\
\end{array}\!
\right]+\eta_e\;\left[\!
\begin{array}{ccc}
\phantom{-} 2 &\;\;\; -1 &\;\;\; -1 \\
-1 &\;\;\; \phantom{-}2 &\;\;\; -1 \\
 -1 &\;\;\; -1 &\;\;\; \phantom{-}2 \\
\end{array}\!
\right],
\end{align}
with $\tilde{\mathbb{S}}_0=\tilde{\mathbb{S}}_e\left(e_1\to - e_1, e_2\to - e_2,\eta_e\to -\eta_o\right)$. Equation~\eqref{md1} shows that for $e_2/ e_1\ll 1$, the charging modes satisfy $\eta_o/\eta_e\ll 1$.
\end{widetext}

\begin{thebibliography}{24}%
\makeatletter
\providecommand \@ifxundefined [1]{%
 \@ifx{#1\undefined}
}%
\providecommand \@ifnum [1]{%
 \ifnum #1\expandafter \@firstoftwo
 \else \expandafter \@secondoftwo
 \fi
}%
\providecommand \@ifx [1]{%
 \ifx #1\expandafter \@firstoftwo
 \else \expandafter \@secondoftwo
 \fi
}%
\providecommand \natexlab [1]{#1}%
\providecommand \enquote  [1]{``#1''}%
\providecommand \bibnamefont  [1]{#1}%
\providecommand \bibfnamefont [1]{#1}%
\providecommand \citenamefont [1]{#1}%
\providecommand \href@noop [0]{\@secondoftwo}%
\providecommand \href [0]{\begingroup \@sanitize@url \@href}%
\providecommand \@href[1]{\@@startlink{#1}\@@href}%
\providecommand \@@href[1]{\endgroup#1\@@endlink}%
\providecommand \@sanitize@url [0]{\catcode `\\12\catcode `\$12\catcode
  `\&12\catcode `\#12\catcode `\^12\catcode `\_12\catcode `\%12\relax}%
\providecommand \@@startlink[1]{}%
\providecommand \@@endlink[0]{}%
\providecommand \url  [0]{\begingroup\@sanitize@url \@url }%
\providecommand \@url [1]{\endgroup\@href {#1}{\urlprefix }}%
\providecommand \urlprefix  [0]{URL }%
\providecommand \Eprint [0]{\href }%
\providecommand \doibase [0]{https://doi.org/}%
\providecommand \selectlanguage [0]{\@gobble}%
\providecommand \bibinfo  [0]{\@secondoftwo}%
\providecommand \bibfield  [0]{\@secondoftwo}%
\providecommand \translation [1]{[#1]}%
\providecommand \BibitemOpen [0]{}%
\providecommand \bibitemStop [0]{}%
\providecommand \bibitemNoStop [0]{.\EOS\space}%
\providecommand \EOS [0]{\spacefactor3000\relax}%
\providecommand \BibitemShut  [1]{\csname bibitem#1\endcsname}%
\let\auto@bib@innerbib\@empty
\bibitem [{\citenamefont {Matveev}(1991)}]{Matveev_1991}%
  \BibitemOpen
  \bibfield  {author} {\bibinfo {author} {\bibfnamefont {K.~A.}\ \bibnamefont
  {Matveev}},\ }\bibfield  {title} {\bibinfo {title} {Quantum fluctuations of
  the charge of a metal particle under the coulomb blockade conditions},\
  }\href@noop {} {\bibfield  {journal} {\bibinfo  {journal} {Sov. Phys. JETP}\
  }\textbf {\bibinfo {volume} {72}},\ \bibinfo {pages} {892} (\bibinfo {year}
  {1991})}\BibitemShut {NoStop}%
\bibitem [{\citenamefont {Averin}\ and\ \citenamefont {Likharev}(1991)}]{AL}%
  \BibitemOpen
  \bibfield  {author} {\bibinfo {author} {\bibfnamefont {D.~V.}\ \bibnamefont
  {Averin}}\ and\ \bibinfo {author} {\bibfnamefont {K.~K.}\ \bibnamefont
  {Likharev}},\ }\bibinfo {title} {Single electronics: A correlated transfer of
  single electrons and cooper pairs in systems of small tunnel junctions},\ in\
  \href@noop {} {\emph {\bibinfo {booktitle} {Mesoscopic Phenomena in
  Solids}}},\ \bibinfo {editor} {edited by\ \bibinfo {editor} {\bibfnamefont
  {B.~L.}\ \bibnamefont {Altshuler}}, \bibinfo {editor} {\bibfnamefont {P.~A.}\
  \bibnamefont {Lee}},\ and\ \bibinfo {editor} {\bibfnamefont {R.~A.}\
  \bibnamefont {Webb}}}\ (\bibinfo  {publisher} {Elsevier},\ \bibinfo {address}
  {Amsterdam},\ \bibinfo {year} {1991})\BibitemShut {NoStop}%
\bibitem [{\citenamefont {Averin}\ and\ \citenamefont
  {Nazarov}(1992)}]{Averin1992}%
  \BibitemOpen
  \bibfield  {author} {\bibinfo {author} {\bibfnamefont {D.~V.}\ \bibnamefont
  {Averin}}\ and\ \bibinfo {author} {\bibfnamefont {Y.~V.}\ \bibnamefont
  {Nazarov}},\ }\bibinfo {title} {Macroscopic quantum tunneling of charge and
  co-tunneling},\ in\ \href@noop {} {\emph {\bibinfo {booktitle} {Single Charge
  Tunneling: Coulomb Blockade Phenomena In Nanostructures}}},\ \bibinfo
  {editor} {edited by\ \bibinfo {editor} {\bibfnamefont {H.}~\bibnamefont
  {Grabert}}\ and\ \bibinfo {editor} {\bibfnamefont {M.~H.}\ \bibnamefont
  {Devoret}}}\ (\bibinfo  {publisher} {Springer US},\ \bibinfo {address}
  {Boston, MA},\ \bibinfo {year} {1992})\BibitemShut {NoStop}%
\bibitem [{\citenamefont {Kastner}(1992)}]{sett}%
  \BibitemOpen
  \bibfield  {author} {\bibinfo {author} {\bibfnamefont {M.~A.}\ \bibnamefont
  {Kastner}},\ }\bibfield  {title} {\bibinfo {title} {The single-electron
  transistor},\ }\href {https://doi.org/10.1103/RevModPhys.64.849} {\bibfield
  {journal} {\bibinfo  {journal} {Rev. Mod. Phys.}\ }\textbf {\bibinfo {volume}
  {64}},\ \bibinfo {pages} {849} (\bibinfo {year} {1992})}\BibitemShut
  {NoStop}%
\bibitem [{\citenamefont {Flensberg}(1993)}]{Flensberg_1993}%
  \BibitemOpen
  \bibfield  {author} {\bibinfo {author} {\bibfnamefont {K.}~\bibnamefont
  {Flensberg}},\ }\bibfield  {title} {\bibinfo {title} {Capacitance and
  conductance of mesoscopic systems connected by quantum point contacts},\
  }\href {https://doi.org/10.1103/PhysRevB.48.11156} {\bibfield  {journal}
  {\bibinfo  {journal} {Phys. Rev. B}\ }\textbf {\bibinfo {volume} {48}},\
  \bibinfo {pages} {11156} (\bibinfo {year} {1993})}\BibitemShut {NoStop}%
\bibitem [{\citenamefont {Matveev}(1995)}]{Matveev1995}%
  \BibitemOpen
  \bibfield  {author} {\bibinfo {author} {\bibfnamefont {K.~A.}\ \bibnamefont
  {Matveev}},\ }\bibfield  {title} {\bibinfo {title} {Coulomb blockade at
  almost perfect transmission},\ }\href
  {https://doi.org/10.1103/PhysRevB.51.1743} {\bibfield  {journal} {\bibinfo
  {journal} {Phys. Rev. B}\ }\textbf {\bibinfo {volume} {51}},\ \bibinfo
  {pages} {1743} (\bibinfo {year} {1995})}\BibitemShut {NoStop}%
\bibitem [{\citenamefont {Furusaki}\ and\ \citenamefont
  {Matveev}(1995)}]{Furusaki1995b}%
  \BibitemOpen
  \bibfield  {author} {\bibinfo {author} {\bibfnamefont {A.}~\bibnamefont
  {Furusaki}}\ and\ \bibinfo {author} {\bibfnamefont {K.~A.}\ \bibnamefont
  {Matveev}},\ }\bibfield  {title} {\bibinfo {title} {Theory of strong
  inelastic cotunneling},\ }\href {https://doi.org/10.1103/PhysRevB.52.16676}
  {\bibfield  {journal} {\bibinfo  {journal} {Phys. Rev. B}\ }\textbf {\bibinfo
  {volume} {52}},\ \bibinfo {pages} {16676} (\bibinfo {year}
  {1995})}\BibitemShut {NoStop}%
\bibitem [{\citenamefont {Jezouin}\ \emph {et~al.}(2016)\citenamefont
  {Jezouin}, \citenamefont {Iftikhar}, \citenamefont {Anthore}, \citenamefont
  {Parmentier}, \citenamefont {Gennser}, \citenamefont {Cavanna}, \citenamefont
  {Ouerghi}, \citenamefont {Levkivskyi}, \citenamefont {Idrisov}, \citenamefont
  {Sukhorukov}, \citenamefont {Glazman},\ and\ \citenamefont
  {Pierre}}]{Pierre_2016}%
  \BibitemOpen
  \bibfield  {author} {\bibinfo {author} {\bibfnamefont {S.}~\bibnamefont
  {Jezouin}}, \bibinfo {author} {\bibfnamefont {Z.}~\bibnamefont {Iftikhar}},
  \bibinfo {author} {\bibfnamefont {A.}~\bibnamefont {Anthore}}, \bibinfo
  {author} {\bibfnamefont {F.~D.}\ \bibnamefont {Parmentier}}, \bibinfo
  {author} {\bibfnamefont {U.}~\bibnamefont {Gennser}}, \bibinfo {author}
  {\bibfnamefont {A.}~\bibnamefont {Cavanna}}, \bibinfo {author} {\bibfnamefont
  {A.}~\bibnamefont {Ouerghi}}, \bibinfo {author} {\bibfnamefont {I.~P.}\
  \bibnamefont {Levkivskyi}}, \bibinfo {author} {\bibfnamefont
  {E.}~\bibnamefont {Idrisov}}, \bibinfo {author} {\bibfnamefont {E.~V.}\
  \bibnamefont {Sukhorukov}}, \bibinfo {author} {\bibfnamefont {L.~I.}\
  \bibnamefont {Glazman}},\ and\ \bibinfo {author} {\bibfnamefont
  {F.}~\bibnamefont {Pierre}},\ }\bibfield  {title} {\bibinfo {title}
  {Controlling charge quantization with quantum fluctuations},\ }\href
  {http://dx.doi.org/10.1038/nature19072} {\bibfield  {journal} {\bibinfo
  {journal} {Nature}\ }\textbf {\bibinfo {volume} {536}},\ \bibinfo {pages}
  {60} (\bibinfo {year} {2016})}\BibitemShut {NoStop}%
\bibitem [{\citenamefont {Idrisov}\ \emph {et~al.}(2017)\citenamefont
  {Idrisov}, \citenamefont {Levkivskyi},\ and\ \citenamefont
  {Sukhorukov}}]{td}%
  \BibitemOpen
  \bibfield  {author} {\bibinfo {author} {\bibfnamefont {E.~G.}\ \bibnamefont
  {Idrisov}}, \bibinfo {author} {\bibfnamefont {I.~P.}\ \bibnamefont
  {Levkivskyi}},\ and\ \bibinfo {author} {\bibfnamefont {E.~V.}\ \bibnamefont
  {Sukhorukov}},\ }\bibfield  {title} {\bibinfo {title} {Thermal decay of
  coulomb blockade oscillations},\ }\href
  {https://doi.org/10.1103/PhysRevB.96.155408} {\bibfield  {journal} {\bibinfo
  {journal} {Phys. Rev. B}\ }\textbf {\bibinfo {volume} {96}},\ \bibinfo
  {pages} {155408} (\bibinfo {year} {2017})}\BibitemShut {NoStop}%
\bibitem [{\citenamefont {Waugh}\ \emph {et~al.}(1995)\citenamefont {Waugh},
  \citenamefont {Berry}, \citenamefont {Mar}, \citenamefont {Westervelt},
  \citenamefont {Campman},\ and\ \citenamefont {Gossard}}]{dd1}%
  \BibitemOpen
  \bibfield  {author} {\bibinfo {author} {\bibfnamefont {F.~R.}\ \bibnamefont
  {Waugh}}, \bibinfo {author} {\bibfnamefont {M.~J.}\ \bibnamefont {Berry}},
  \bibinfo {author} {\bibfnamefont {D.~J.}\ \bibnamefont {Mar}}, \bibinfo
  {author} {\bibfnamefont {R.~M.}\ \bibnamefont {Westervelt}}, \bibinfo
  {author} {\bibfnamefont {K.~L.}\ \bibnamefont {Campman}},\ and\ \bibinfo
  {author} {\bibfnamefont {A.~C.}\ \bibnamefont {Gossard}},\ }\bibfield
  {title} {\bibinfo {title} {Single-electron charging in double and triple
  quantum dots with tunable coupling},\ }\href
  {https://doi.org/10.1103/PhysRevLett.75.705} {\bibfield  {journal} {\bibinfo
  {journal} {Phys. Rev. Lett.}\ }\textbf {\bibinfo {volume} {75}},\ \bibinfo
  {pages} {705} (\bibinfo {year} {1995})}\BibitemShut {NoStop}%
\bibitem [{\citenamefont {Matveev}\ \emph {et~al.}(1996)\citenamefont
  {Matveev}, \citenamefont {Glazman},\ and\ \citenamefont {Baranger}}]{kamdd}%
  \BibitemOpen
  \bibfield  {author} {\bibinfo {author} {\bibfnamefont {K.~A.}\ \bibnamefont
  {Matveev}}, \bibinfo {author} {\bibfnamefont {L.~I.}\ \bibnamefont
  {Glazman}},\ and\ \bibinfo {author} {\bibfnamefont {H.~U.}\ \bibnamefont
  {Baranger}},\ }\bibfield  {title} {\bibinfo {title} {Coulomb blockade of
  tunneling through a double quantum dot},\ }\href
  {https://doi.org/10.1103/PhysRevB.54.5637} {\bibfield  {journal} {\bibinfo
  {journal} {Phys. Rev. B}\ }\textbf {\bibinfo {volume} {54}},\ \bibinfo
  {pages} {5637} (\bibinfo {year} {1996})}\BibitemShut {NoStop}%
\bibitem [{\citenamefont {Golden}\ and\ \citenamefont
  {Halperin}(1996)}]{kamdd1}%
  \BibitemOpen
  \bibfield  {author} {\bibinfo {author} {\bibfnamefont {J.~M.}\ \bibnamefont
  {Golden}}\ and\ \bibinfo {author} {\bibfnamefont {B.~I.}\ \bibnamefont
  {Halperin}},\ }\bibfield  {title} {\bibinfo {title} {Relation between barrier
  conductance and coulomb blockade peak splitting for tunnel-coupled quantum
  dots},\ }\href {https://doi.org/10.1103/PhysRevB.53.3893} {\bibfield
  {journal} {\bibinfo  {journal} {Phys. Rev. B}\ }\textbf {\bibinfo {volume}
  {53}},\ \bibinfo {pages} {3893} (\bibinfo {year} {1996})}\BibitemShut
  {NoStop}%
\bibitem [{\citenamefont {van~der Wiel}\ \emph {et~al.}(2002)\citenamefont
  {van~der Wiel}, \citenamefont {De~Franceschi}, \citenamefont {Elzerman},
  \citenamefont {Fujisawa}, \citenamefont {Tarucha},\ and\ \citenamefont
  {Kouwenhoven}}]{koven}%
  \BibitemOpen
  \bibfield  {author} {\bibinfo {author} {\bibfnamefont {W.~G.}\ \bibnamefont
  {van~der Wiel}}, \bibinfo {author} {\bibfnamefont {S.}~\bibnamefont
  {De~Franceschi}}, \bibinfo {author} {\bibfnamefont {J.~M.}\ \bibnamefont
  {Elzerman}}, \bibinfo {author} {\bibfnamefont {T.}~\bibnamefont {Fujisawa}},
  \bibinfo {author} {\bibfnamefont {S.}~\bibnamefont {Tarucha}},\ and\ \bibinfo
  {author} {\bibfnamefont {L.~P.}\ \bibnamefont {Kouwenhoven}},\ }\bibfield
  {title} {\bibinfo {title} {Electron transport through double quantum dots},\
  }\href {https://doi.org/10.1103/RevModPhys.75.1} {\bibfield  {journal}
  {\bibinfo  {journal} {Rev. Mod. Phys.}\ }\textbf {\bibinfo {volume} {75}},\
  \bibinfo {pages} {1} (\bibinfo {year} {2002})}\BibitemShut {NoStop}%
\bibitem [{\citenamefont {Pouse}\ \emph {et~al.}(2023)\citenamefont {Pouse},
  \citenamefont {Peeters}, \citenamefont {Hsueh}, \citenamefont {Gennser},
  \citenamefont {Cavanna}, \citenamefont {Kastner}, \citenamefont {Mitchell},\
  and\ \citenamefont {Goldhaber-Gordon}}]{ccd}%
  \BibitemOpen
  \bibfield  {author} {\bibinfo {author} {\bibfnamefont {W.}~\bibnamefont
  {Pouse}}, \bibinfo {author} {\bibfnamefont {L.}~\bibnamefont {Peeters}},
  \bibinfo {author} {\bibfnamefont {C.~L.}\ \bibnamefont {Hsueh}}, \bibinfo
  {author} {\bibfnamefont {U.}~\bibnamefont {Gennser}}, \bibinfo {author}
  {\bibfnamefont {A.}~\bibnamefont {Cavanna}}, \bibinfo {author} {\bibfnamefont
  {M.~A.}\ \bibnamefont {Kastner}}, \bibinfo {author} {\bibfnamefont {A.~K.}\
  \bibnamefont {Mitchell}},\ and\ \bibinfo {author} {\bibfnamefont
  {D.}~\bibnamefont {Goldhaber-Gordon}},\ }\bibfield  {title} {\bibinfo {title}
  {Quantum simulation of an exotic quantum critical point in a two-site charge
  kondo circuit},\ }\href {https://doi.org/10.1038/s41567-022-01905-4}
  {\bibfield  {journal} {\bibinfo  {journal} {Nat. Phys.}\ }\textbf {\bibinfo
  {volume} {19}},\ \bibinfo {pages} {492} (\bibinfo {year} {2023})}\BibitemShut
  {NoStop}%
\bibitem [{\citenamefont {Karki}\ \emph {et~al.}(2023)\citenamefont {Karki},
  \citenamefont {Boulat}, \citenamefont {Pouse}, \citenamefont
  {Goldhaber-Gordon}, \citenamefont {Mitchell},\ and\ \citenamefont
  {Mora}}]{KBM1}%
  \BibitemOpen
  \bibfield  {author} {\bibinfo {author} {\bibfnamefont {D.~B.}\ \bibnamefont
  {Karki}}, \bibinfo {author} {\bibfnamefont {E.}~\bibnamefont {Boulat}},
  \bibinfo {author} {\bibfnamefont {W.}~\bibnamefont {Pouse}}, \bibinfo
  {author} {\bibfnamefont {D.}~\bibnamefont {Goldhaber-Gordon}}, \bibinfo
  {author} {\bibfnamefont {A.~K.}\ \bibnamefont {Mitchell}},\ and\ \bibinfo
  {author} {\bibfnamefont {C.}~\bibnamefont {Mora}},\ }\bibfield  {title}
  {\bibinfo {title} {$\mathbb{Z}_3$ parafermion in the double charge kondo
  model},\ }\href {https://doi.org/10.1103/PhysRevLett.130.146201} {\bibfield
  {journal} {\bibinfo  {journal} {Phys. Rev. Lett.}\ }\textbf {\bibinfo
  {volume} {130}},\ \bibinfo {pages} {146201} (\bibinfo {year}
  {2023})}\BibitemShut {NoStop}%
\bibitem [{\citenamefont {Karki}\ \emph {et~al.}(2022)\citenamefont {Karki},
  \citenamefont {Boulat},\ and\ \citenamefont {Mora}}]{KBM}%
  \BibitemOpen
  \bibfield  {author} {\bibinfo {author} {\bibfnamefont {D.~B.}\ \bibnamefont
  {Karki}}, \bibinfo {author} {\bibfnamefont {E.}~\bibnamefont {Boulat}},\ and\
  \bibinfo {author} {\bibfnamefont {C.}~\bibnamefont {Mora}},\ }\bibfield
  {title} {\bibinfo {title} {Double-charge quantum island in the quasiballistic
  regime},\ }\href {https://doi.org/10.1103/PhysRevB.105.245418} {\bibfield
  {journal} {\bibinfo  {journal} {Phys. Rev. B}\ }\textbf {\bibinfo {volume}
  {105}},\ \bibinfo {pages} {245418} (\bibinfo {year} {2022})}\BibitemShut
  {NoStop}%
\bibitem [{\citenamefont {Wen}(1992)}]{Wen_1990}%
  \BibitemOpen
  \bibfield  {author} {\bibinfo {author} {\bibfnamefont {X.-G.}\ \bibnamefont
  {Wen}},\ }\bibfield  {title} {\bibinfo {title} {Theory of the edge states in
  fractional quantum hall effects},\ }\href
  {https://doi.org/10.1142/S0217979292000840} {\bibfield  {journal} {\bibinfo
  {journal} {Int. J. Mod. Phys. B}\ }\textbf {\bibinfo {volume} {06}},\
  \bibinfo {pages} {1711} (\bibinfo {year} {1992})}\BibitemShut {NoStop}%
\bibitem [{Note1()}]{Note1}%
  \BibitemOpen
  \bibinfo {note} {We assume the same Fermi velocity $v_{\protect \rm F}$ for
  each edge channel.}\BibitemShut {Stop}%
\bibitem [{\citenamefont {Slobodeniuk}\ \emph {et~al.}(2013)\citenamefont
  {Slobodeniuk}, \citenamefont {Levkivskyi},\ and\ \citenamefont
  {Sukhorukov}}]{Slobodeniuk_2013}%
  \BibitemOpen
  \bibfield  {author} {\bibinfo {author} {\bibfnamefont {A.~O.}\ \bibnamefont
  {Slobodeniuk}}, \bibinfo {author} {\bibfnamefont {I.~P.}\ \bibnamefont
  {Levkivskyi}},\ and\ \bibinfo {author} {\bibfnamefont {E.~V.}\ \bibnamefont
  {Sukhorukov}},\ }\bibfield  {title} {\bibinfo {title} {Equilibration of
  quantum hall edge states by an ohmic contact},\ }\href
  {https://doi.org/10.1103/PhysRevB.88.165307} {\bibfield  {journal} {\bibinfo
  {journal} {Phys. Rev. B}\ }\textbf {\bibinfo {volume} {88}},\ \bibinfo
  {pages} {165307} (\bibinfo {year} {2013})}\BibitemShut {NoStop}%
\bibitem [{\citenamefont {Sukhorukov}(2016)}]{Sukhorukov_2016}%
  \BibitemOpen
  \bibfield  {author} {\bibinfo {author} {\bibfnamefont {E.~V.}\ \bibnamefont
  {Sukhorukov}},\ }\bibfield  {title} {\bibinfo {title} {Scattering theory
  approach to bosonization of non-equilibrium mesoscopic systems},\ }\href
  {https://doi.org/https://doi.org/10.1016/j.physe.2015.11.018} {\bibfield
  {journal} {\bibinfo  {journal} {Phys. E}\ }\textbf {\bibinfo {volume} {77}},\
  \bibinfo {pages} {191} (\bibinfo {year} {2016})}\BibitemShut {NoStop}%
\bibitem [{\citenamefont {Morel}\ \emph {et~al.}(2022)\citenamefont {Morel},
  \citenamefont {Lee}, \citenamefont {Sim},\ and\ \citenamefont
  {Mora}}]{morel2021}%
  \BibitemOpen
  \bibfield  {author} {\bibinfo {author} {\bibfnamefont {T.}~\bibnamefont
  {Morel}}, \bibinfo {author} {\bibfnamefont {J.-Y.~M.}\ \bibnamefont {Lee}},
  \bibinfo {author} {\bibfnamefont {H.-S.}\ \bibnamefont {Sim}},\ and\ \bibinfo
  {author} {\bibfnamefont {C.}~\bibnamefont {Mora}},\ }\bibfield  {title}
  {\bibinfo {title} {Fractionalization and anyonic statistics in the integer
  quantum hall collider},\ }\href {https://doi.org/10.1103/PhysRevB.105.075433}
  {\bibfield  {journal} {\bibinfo  {journal} {Phys. Rev. B}\ }\textbf {\bibinfo
  {volume} {105}},\ \bibinfo {pages} {075433} (\bibinfo {year}
  {2022})}\BibitemShut {NoStop}%
\bibitem [{Note2()}]{Note2}%
  \BibitemOpen
  \bibinfo {note} {Linear conductance in yet another regime, where temperature
  and the strengths of charging modes are identical, of the considered device
  can be evaluated analytically.}\BibitemShut {Stop}%
\bibitem [{\citenamefont {Falci}\ \emph {et~al.}(1991)\citenamefont {Falci},
  \citenamefont {Bubanja},\ and\ \citenamefont {Sch{\"o}n}}]{sochin}%
  \BibitemOpen
  \bibfield  {author} {\bibinfo {author} {\bibfnamefont {G.}~\bibnamefont
  {Falci}}, \bibinfo {author} {\bibfnamefont {V.}~\bibnamefont {Bubanja}},\
  and\ \bibinfo {author} {\bibfnamefont {G.}~\bibnamefont {Sch{\"o}n}},\
  }\bibfield  {title} {\bibinfo {title} {Quasiparticle and cooper pair
  tenneling in small capacitance josephson junctions},\ }\href
  {https://doi.org/10.1007/BF01307643} {\bibfield  {journal} {\bibinfo
  {journal} {Z. Phys. B}\ }\textbf {\bibinfo {volume} {85}},\ \bibinfo {pages}
  {451} (\bibinfo {year} {1991})}\BibitemShut {NoStop}%
\bibitem [{\citenamefont {Nguyen}\ and\ \citenamefont {Kiselev}(2018)}]{mkk}%
  \BibitemOpen
  \bibfield  {author} {\bibinfo {author} {\bibfnamefont {T.~K.~T.}\
  \bibnamefont {Nguyen}}\ and\ \bibinfo {author} {\bibfnamefont {M.~N.}\
  \bibnamefont {Kiselev}},\ }\bibfield  {title} {\bibinfo {title} {Seebeck
  effect on a weak link between fermi and non-fermi liquids},\ }\href
  {https://doi.org/10.1103/PhysRevB.97.085403} {\bibfield  {journal} {\bibinfo
  {journal} {Phys. Rev. B}\ }\textbf {\bibinfo {volume} {97}},\ \bibinfo
  {pages} {085403} (\bibinfo {year} {2018})}\BibitemShut {NoStop}%
\end{thebibliography}
%
\end{document}